\documentclass[prd,superscriptaddress,showkeys,showpacs,onecolumn,showkeys]{revtex4}
\usepackage{eurosym}
\usepackage{amsfonts}
\usepackage{array}
\usepackage{amsthm}
\usepackage{bm}
\usepackage{palatino}
\usepackage{mathpazo}
\usepackage{amssymb}
\usepackage{eurosym}
\usepackage{amsmath}
\usepackage{epsfig}
\usepackage{graphics}
\usepackage{changes}
\usepackage{color}
\usepackage{graphicx}
\usepackage[colorlinks=true,
            linkcolor=blue,
           urlcolor=blue,
           citecolor=blue]{hyperref}

\def\be{\begin{equation}}
\def\ee{\end{equation}}
\def\beq{\begin{eqnarray}}
\def\eeq{\end{eqnarray}}

\def\bes{\begin{eqnarray}}
\def\ees{\end{eqnarray}}

\begin{document}
\title{Tunneling of Massive Vector Particles under the Influence of Quantum Gravity}
\author{Wajiha Javed}
\email{wajiha.javed@ue.edu.pk; wajihajaved84@yahoo.com} 
\affiliation{Division of Science and Technology, University of Education, Township-Lahore, Pakistan}

\author{Riasat Ali}
\email{riasatyasin@gmail.com}
\affiliation{Department of Mathematics, Government College,
University Faisalabad Layyah Campus, Layyah-31200, Pakistan}

\author{Rimsha Babar}
\email{rimsha.babar10@gmail.com} 
\affiliation{Division of Science and Technology, University of Education, Township-Lahore, Pakistan}

\author{Ali \"{O}vg\"{u}n}
\email{ali.ovgun@pucv.cl}
\homepage[]{https://www.aovgun.com}
\affiliation{Instituto de F\'{\i}sica, Pontificia Universidad Cat\'olica de Valpara\'{\i}%
so, Casilla 4950, Valpara\'{\i}so, Chile}
\affiliation{Physics Department, Faculty of Arts and Sciences, Eastern Mediterranean
University, Famagusta, North Cyprus, via Mersin 10, Turkey}

\begin{abstract}
This paper is devoted to investigate charged vector particles tunneling via horizons of a 
pair of accelerating rotating charged NUT black hole under the influence of 
quantum gravitational effects. For this purpose,
we use the modified Proca equation incorporating
generalized uncertainty principle. Using the WKB approximation to the field equation,
we obtain a modified tunneling rate and the corresponding corrected Hawking temperature for this
black hole. Moreover, we analyze the graphical behavior
of corrected Hawking temperature $T'_{H}$ with respect to the event horizon for the given black hole.
By considering quantum gravitational effects on Hawking temperatures, we discuss the stability analysis of
this black hole. For a pair of black holes, the temperature $T'_{H}$ increases with
the increase in rotation parameters $a$ and $\omega$, correction parameter
$\beta$, black hole acceleration $\alpha$ and arbitrary parameter $k$ and decreases
with the increase in electric $e$ and magnetic charges $g$.
\end{abstract}



\date{\today}
\maketitle
\section{Introduction}

A black hole (BH) is a physical object that has the ability to absorb all types
of energy from the surrounding due to its strong gravitational pull. According to the theory of
general relativity, a BH attracts all types of particles that interact with event horizon.
Hawking $(1974)$ described that, a BH acts as a black body and emit particles in form of radiation
through its horizon by considering quantum effects in the background of curved spacetime,
these radiation are known as \textit{Hawking radiation} \cite{B17} and possesses a
particular temperature which is called \textit{Hawking temperature}.
In order to study the Hawking radiation phenomenon, the \textit{quantum tunneling} is
one of the best technique \cite{[1]}-\cite{d1}. This technique depends on electron-positron
pair-creation, which needs an electric field. It could be a well-known fact that as particles
overcome the horizon, the energy changes the sign, so that a pair created just inside/outside
the horizon can materialize with zero total energy, while
one member of the pair has ability to tunnel to the opposite side of horizon \cite{[5]}.

According to the tunneling phenomenon, particles are permitted to
follow the classically forbidden trajectories, by starting from outside the
horizon to infinity \cite{[6],[7]}. The BH evaporation can be studied
with the discharge of quantum particles in form of Hawking radiation which enables BH to lose its mass.
When a BH loses more matter instead of increasing through different means
then it disseminate, shrink and eventually disappears. This phenomenon
causes a change in thermodynamical characteristic of a BH, i.e.,
charge, mass and angular momentum. These particles follow the frame of
outgoing/ingoing radial null geodesics. For the outgoing geodesics, the particles
must be imaginary although for ingoing geodesics, they are considered to be real.
It is due to the fact that just a real particle that has a speed not exactly or equivalent to the
speed of light can exist inside the horizon.

For calculating the imaginary part of the classical action there are two main
approaches, i.e., \textit{null geodesic technique} and \textit{Hamilton-Jacobi strategy}.
The first one proposed by Parikh and his colleagues \cite{[5]}
and second one introduced by Srinivasan and Padmanabhan \cite{[6]}.
To construct a relation between classical and quantum theory, an approximation is
known as \textit{WKB approximation} utilized by Wentzel, Kramers and Brillouin (WKB) \cite{[8]}. Various authors studied tunneling of vector particles, bosonic particles,  spin-2, spin-3/2 and fermionic particles  to obtain the Hawking temperature for different BHs and wormholes \cite{R55,Sakalli:2015taa,Sakalli:2014sea,Sakalli:2017ewb,Sakalli:2015mka,Sakalli:2015nza,Sakalli:2016cbo,Ovgun:2015jna,Ovgun:2017hje,Jusufi:2017vhz,Kuang:2017sqa,Kuang:2018goo,Javed:2018msn,Gonzalez:2017zdz,Ovgun:2016roz,Jusufi:2016hcn,deyou1,deyou2,aa19,aa20,aa21,aw1,za4,za5,za6,za7,za2,za3,Akhmedova:2010zz,deGill:2010nb,Zhu:2009wa,Akhmedova:2008dz,Akhmedova:2008au,Akhmedov:2006un,Kanzi:2019gtu,Rizwan:2018gpl,Gecim:2018sji,Meitei:2018mgo,Chen:2014eka,Zeng:2009zzd,Modak:2008tg,Li:2008ws,Li:2007zzg,Jiang:2007pn,He:2007zz,He:2007zzi,Li:2006rg,R66,R7,R8,R44}.

By incorporating generalized uncertainty principle (GUP) effects, it is conceivable to discuss quantum corrected thermodynamical properties of BH \cite{5}.
The GUP gives high-energy remedies to BH thermodynamics, which leads to the possibility of a minimal
length in quantum gravity theory. The idea of GUP has been utilized for various BHs.
The tunneling process for Kerr, Kerr-Newman and Reissner-Nordstr\"{o}m (RN) BHs \cite{Rv1} provides
important contribution towards the BH physics. Jiang \cite{R1} calculated
the tunneling of Dirac particles and analyzed the
Hawking radiation spectrum for black ring. 
Jian and Bing-Bing \cite{R3} studied the fermion particles tunneling from uncharged and
charged BHs. Yale \cite{R4} studied the scalar, fermion and boson particles
tunneling from BHs without back-reaction effects, they also analyzed the exact Hawking temperature
for these particles. Sharif and Javed \cite{a4} studied the Hawking radiation
through quantum tunneling process for various types of BHs and also derived
the tunneling probabilities as well as their corresponding Hawking temperatures.
The same authors \cite{R5} calculated the tunneling behavior
of fermion particles from traversable wormholes. They also discussed the tunneling
behavior of fermion particles for charged accelerating rotating BHs
associated with $NUT$ parameter \cite{R6}.

$\ddot{\textmd{O}}$vg$\ddot{\textmd{u}}$n et al. \cite{B56} calculated the tunneling
rate of charged massive bosons for various types of BHs surrounded by the perfect fluid in Rastall theory.
Anacleto et al. \cite{R15} studied the finite quantum corrected entropy from non-commutative
acoustic BHs. Li and Zu \cite{R19} analyzed the GUP effects by utilizing Klein-Gordon
equation and also studied corrected temperature for Gibbon-Maeda-Dilation BH.
Anacleto et al. \cite{R16} investigated the quantum corrected entropy by using tunneling
method with GUP in self-dual BH. \"{O}vg\"{u}n and Jusufi \cite{R20} studied the
spin-$1$ vector particles tunneling for charged non-commutative BHs and its
GUP-corrected thermodynamical properties. Singh et al. \cite{R21} analyzed
particles radiation for Kerr-Newman BH by using quantum tunneling phenomenon.
Sakalli et al. \cite{Rv2} investigated the scalar particles tunneling from acoustic
BHs with a rotation parameter by applying $WKB$ approximation and
Klein-Gordon equation with GUP to obtain Hawking temperature of massive particles.

Javed et al. \cite{A1} analyzed charged vector particles tunneling phenomenon for
a pair of accelerating and rotating as well as 5D gauged super-gravity BHs and
investigated the corresponding Hawking temperatures. In the continuation of their
work, we study the Hawking radiation phenomenon by considering quantum corrections of
boson particles through the event horizon of accelerating and rotating BH
with $NUT$ parameter.
We also discuss the graphical behavior of
corrected temperature with respect to horizon for given BH and analyzed the
effects of different parameters on temperature and visualize the stable
and unstable states of BH. The main motivation of this paper is to study the charged accelerating and rotating
black hole with $NUT$ parameters under the influence of quantum gravity effects
and to check how the results reduce to the previous results when we neglect the quantum gravity effects as well as other parameters.
The organization of this paper is as follows: In section \textbf{2}, we study
the quantum corrected tunneling rate and corrected Hawking temperature for charged BH
solution having acceleration, rotation and NUT parameter. Moreover, we analyze
the graphical behavior of corrected temperature with respect to the event
horizon of BH and also discuss the effects of different parameters on
temperature. Section \textbf{3} is based on the conclusion and discussion of the mathematical results for these BHs.

\section{Pair of Charged Accelerated Black Holes Involving Rotation and NUT Parameters}

Universally, the $NUT$ parameter is associated with the twisting behavior of the
surrounding spacetime or with the gravito-magnetic monopole parameter of the central mass. Moreover, its precise
physical significance might not be investigated. Later, the higher dimensional
origin of the Kerr-NUT-(anti) de-Sitter BH
and its physical significance is studied \cite{R22, R24}. For a
BH, the apparent dominance of the $NUT$ parameter on the rotation parameter
takes off the metric free of curvature singularity and the corresponding
result is known as $NUT$-like result. When the rotation parameter
commands the $NUT$ parameter, the solution is Kerr-like and a ring
curvature singularity occurs. This type of singularity is independent from existence of the
cosmological constant.

The exact understanding of the NUT parameter is conceivable when a static Schwarzschild
mass is submerged in a stationary source-free electromagnetic universe \cite{R25}.
Moreover, the $NUT$ parameter is associated with the representation of
twisting behavior of the electromagnetic universe allowing off the
Schwarzschild central mass. When electric and magnetic
fields vanish, the NUT parameter represents the twist of the vacuum spacetime \cite{R6}.
Consequently, for creation of $NUT$ parameter, the bending of the environmental space couples with the mass of
background field.

The line-element of this BH can be defined as follows \cite{R27}
\begin{eqnarray}\nonumber
ds^{2}&=&-\frac{1}{\Omega^{2}}\Big[-\frac{Q}{\rho^{2}}\Big(
(4l\sin^{2}\frac{\theta}{2}+a\sin^{2}\theta)d\phi-dt\Big)^{2}
-\frac{\rho^{2}}{Q}dr^{2}\nonumber\\&& +\frac{\tilde{P}}{\rho^{2}}
\Big(((l+a)^{2}+r^{2})d\phi-a dt\Big)^{2}
-\sin^{2}\theta\frac{\rho^{2}}{\tilde{P}} d\theta^{2}\Big],\label{1}
\end{eqnarray}
where
\begin{eqnarray}
\Omega&=&-(a\cos\theta+l)\frac{\alpha}{\omega}r+1
,~~~\rho^{2}=(a\cos\theta+l)^{2}+r^{2},\nonumber\\
Q&=&\left[(\tilde{e}^{2}+\omega^{2}\tilde{k}+\tilde{g}^{2})(1+2l\alpha
\frac{r}{\omega})-2rM-\frac{\omega^{2}\tilde{k}r^{2}}{l^{2}-a^{2}}\right]
\nonumber\\
&\times&\left[1-\frac{l-a}{\omega}\alpha r\right]\left[1-\frac{l+a}{\omega}r\alpha\right]
,\nonumber\\
\tilde{P}&=&-\sin^{2}\theta(a_{4}\cos^{2}\theta+a_{3}\cos\theta-1)=\sin^{2}\theta P,\nonumber\\
a_{3}&=&\frac{\alpha a}{\omega}2M-4\frac{\alpha^{2}al}{\omega^{2}}
(\tilde{g}^{2}+\tilde{e}^{2}+\omega^{2}\tilde{k}),\nonumber\\
a_{4}&=&-\frac{\alpha^{2}a^{2}}{\omega^{2}}(\tilde{g}^{2}+\tilde{e}^{2}+\omega^{2}\tilde{k})
,\nonumber
\end{eqnarray}
where $M$, $g$ and $e$ represent mass,
magnetic and electric charges, respectively. While $l$,
$\omega$ and $\alpha$ are $NUT$
parameter, rotation and acceleration of source, respectively. Moreover, $a$ is
rotation parameter (parameter of Kerr-like) and $\tilde{k}$ is defined as
\begin{equation}
\left(\frac{\omega^{2}}{l^{2}-a^{2}}-3\alpha^{2}l^{2}\right)
\tilde{k}=3\frac{\alpha^{2}l^{2}}{\omega^{2}}(\tilde{e}^{2}+\tilde{g}^{2})
-2\frac{\alpha
l}
{\omega}M-1.\nonumber
\end{equation}
We observe that $\omega$ depends on $a$ and $l$. The parameter $\alpha$ represents
the bending behavior of BHs and relative
to the rotation parameter ($\omega$). The parameters $\tilde{e}$, $\tilde{g}$, $M$,
$\omega$, $\tilde{k}$ and $\alpha$ are independent. As
$\alpha=0$, then the line-element (\ref{1}) implies to the
Kerr-Newman $NUT$ results. For $l=0$, the line-element (\ref{1})
gives the BHs with rotating and charged  parameters. When
$\tilde{g}$ and $\tilde{e}$ approaches to zero then we get a
Schwarzschild BH and if $a$ and $l$ approaches to zero, then it implies to the
C-metric.

The line-element (\ref{1}) can be rewritten in the following form
\begin{equation}
ds^{2}=-Z(r, \theta)dt^{2}+\frac{dr^{2}}{B(r, \theta)}
+C(r, \theta)d\theta^{2}+D(r, \theta)d\phi^{2}
-2F(r, \theta)d\phi dt,
\end{equation}
where the metric functions $Z$, $B$, $C$,
$D$ and $F$ are given as follows
\begin{eqnarray*}
Z(r, \theta)&=&\frac{-Pa^2\sin^2\theta+Q}{\Omega^2\rho^2},\quad
B(r, \theta)=\frac{\Omega^2 Q}{\rho^2},\quad C(r, \theta)=\frac{\rho^2}{P\Omega^2},\\
D(r, \theta)&=&\frac{1}{\rho^2\Omega^2}\Big(sin^2\theta
Q((a+l)^2+r^2)^2-Q(4l\sin^2\frac{\theta}{2}+a\sin^2\theta)^2\Big),\\
F(r, \theta)&=&\frac{1}{\rho^2\Omega^2}\Big(Pa sin^2\theta
((l+a)^2+r^2)-Q(4l\sin^2\frac{\theta}{2}+a\sin^2\theta)\Big).
\end{eqnarray*}
The electromagnetic potential for these pair of BHs is defined as
\begin{eqnarray}
A&=&\frac{1}{a((acos\theta+l)^2+r^2)}\Big[\tilde{g}\Big(-l-a\cos\theta\Big(a
dt-\Big(r^2+(a+l)^2\Big)d\phi\nonumber\\&+&r\tilde{e}\Big(d\phi((a+l)-adt
+(2al\cos\theta+a^2\cos^2\theta+l^2))\Big)
\Big]\nonumber.
\end{eqnarray}
The horizons are obtained from
$B(r, \theta)=\frac{\Omega^2 Q}{\rho^2}=0$, which implies that
$\Omega$ is not zero, thus $Q$ is zero, so $r$ has real roots, i.e.,
\begin{eqnarray}
&&r_{\alpha1}=\frac{\omega}{\alpha(l+a)},\quad
r_{\alpha2}=-\frac{\omega}{(a-l)\alpha},\quad
r_{\pm}=\frac{l^2-a^2}{\omega^2\tilde{k}}\Big[\Big((\tilde{g}^2
+\tilde{e}^2+\omega^2\tilde{k})\frac{\alpha
l}{\omega}\nonumber\\
&&-M\Big)\pm\sqrt{\Big((\tilde{g}^2+\tilde{e}^2+\omega^2\tilde{k})\frac{\alpha
l}{\omega}-M\Big)^2+(\tilde{g}^2+\tilde{e}^2+\omega^2\tilde{k})\frac{\omega^2\tilde{k}}{l^2-\alpha^2
}} \Big].
\end{eqnarray}
Here, $r_{\alpha1}$ and $r_{\alpha2}$ represent the acceleration horizons and
$r_\pm$ denote the inner and outer horizons such that
\begin{equation*}{\Big((\tilde{g}^2+\tilde{e}^2+\omega^2\tilde{k})\frac{\alpha
l}{\omega}-M\Big)^2-(\tilde{g}^2+\tilde{e}^2+\omega^2\tilde{k})\frac{\omega^2\tilde{k}}{a^2
-l^2}}>0.
\end{equation*}
The angular velocity of BH at outer horizon is given as follows
\begin{equation}
{{\Omega}_H}= \frac{a}{(a+l)^2+r^2_+}.
\end{equation}
In order to study the corrected tunneling rate for boson
particles through the pair of BH horizon, we consider Lagrangian equation
with quantum gravity effects. By considering a spacetime with
potential due to both electric and magnetic field, the behavior of massive (spin-1) boson
is described by the wave equation. The
GUP modified Lagrangian equation with vector field $\psi_{\mu}$ for bosons particles can be
expressed as \cite{R13,E25}
\begin{equation}
\pounds^{GUP}=-\frac{1}{2}(\mathfrak{D}_{\mu}\psi_{\nu}
-\mathfrak{D}_{\nu}\psi_{\mu})
(\mathfrak{D}^{\mu}\psi^{\nu}-\mathfrak{D}^{\nu}\psi^{\mu})-\frac{\iota}{\hbar}eF^{\nu\mu}\psi_{\mu}\psi_{\nu}-
\frac{m^2}{\hbar^{2}}\psi_{\mu}\psi^{\nu}.
\end{equation}
The modified wave equation for massive bosons can be defined as follows
\begin{eqnarray}
&&\partial_{\mu}(\sqrt{-g}\psi^{\nu\mu})+\sqrt{-g}
\frac{m^{2}}{\hbar^{2}}\psi^{\nu}+\sqrt{-g}\frac{\iota}{\hbar}e A_{\mu}\psi^{\nu\mu}+\sqrt{-g}
\frac{\iota}{\hbar}eF^{\nu\mu}\psi_{\mu}+\beta \hbar^{2}
\partial_{0}\partial_{0}\partial_{0}\nonumber\\&&(\sqrt{-g}g^{00}\psi^{0\nu})-
\beta \hbar^{2}
\partial_{i}\partial_{i}\partial_{i}(\sqrt{-g}g^{ii}\psi^{i\nu})=0,\label{2}
\end{eqnarray}
where $\mid g\mid$, $m$ and $\psi^{\mu\nu}$ are the matrix coefficients, mass
of a particle and anti-symmetric tensor respectively. The anti-symmetric tensor can be defined as
\begin{eqnarray}
\psi_{\nu\mu}&=&(1-\beta \hbar^{2}\partial^{2}_{\nu})\partial_{\nu}\psi_{\mu}-
(1-\beta \hbar^{2}\partial^{2}_{\mu})\partial_{\mu}\psi_{\nu}+
(1-\beta \hbar^{2}\partial^{2}_{\nu})\frac{\iota}{\hbar}e A_{\nu}\psi_{\mu}\nonumber\\
&-&(1-\beta \hbar^{2}\partial^{2}_{\mu})\frac{\iota}{\hbar}e A_{\mu}\psi_{\nu},~~\textmd{and}
\end{eqnarray}
\begin{equation}
F_{\mu\nu}=\nabla_{\mu}A_{\nu}
-\nabla_{\nu}A_{\mu},\nonumber~\textmd{where}~
\nabla_{0}=(1+\beta \hbar^{2}g^{00}\nabla_{0}^2)\nabla_{0},~\nabla_{0}=(1-\beta \hbar^{2}g^{ii}\nabla_{i}^2)\nabla_{i},
\end{equation}
where $\beta$ is the representation of the quantum gravity effect in terms of correction parameter,
if $\beta=0$ and $A=0$, then the above wave equation reduces to the
expressions given in Ref.\cite{A1} and Ref.\cite{B56}, respectively,
if both $\beta$ and $A$ are zero then this wave equation reduces to
the wave equation given in Ref.\cite{R21}. While, $A_\mu$ is considered as the BH
electromagnetic potential, $e$ denotes the bosons particles charge and $\nabla_{\mu}$ is
representing covariant derivative. In the equation of wave motion
the $+ive$ and $-ive$ bosons have alike behavior, the quantum tunneling phenomena
for both type of particles is also similar.
For $+ive$ field, the values of $\psi^{\mu}$ and
$\psi^{\nu\mu}$ are obtained as follows
\begin{eqnarray}
&&\psi^{0}=\frac{-D\psi_{0}-F\psi_{3}}{ZD+F^{2}},~~~ \psi^{1}=B
\psi_{1},~~~ \psi^{2}=C^{-1}
\psi_{2},~~~\psi^{3}=\frac{-F\psi_{0}+Z\psi_{3}}{ZD+F^{2}},\nonumber\\
&&\psi^{01}=\frac{-BD\psi_{01}-BF\psi_{13}}{ZD+F^{2}}
,~~~\psi^{02}=\frac{-D\psi_{02}-F\psi_{23}}{C(
ZD+F^{2})},~~~\psi^{03}=\frac{-\psi_{03}}{ZD+F^{2}},\nonumber\\
&&\psi^{12}=BC^{-1}\psi_{12},~~~\psi^{13}=\frac{B(Z\psi_{13}-F\psi_{01})}{ZD+F^{2}
},~~~\psi^{23}=\frac{B\psi_{23}-F\psi_{02}}{C(ZD+F^{2})}.\nonumber
\end{eqnarray}
Utilizing WKB approximation \cite{R29}
\begin{equation}
\psi_{\nu}=c_{\nu}\exp\left[\frac{\iota}{\hbar}\tilde{I}_{o}\alpha_{n}+
\Sigma \hbar^{n}\tilde{I}_{n}\alpha_{n}\right],
\end{equation}
where $\alpha_{n}=(t, r, \theta, \phi)$ (for $n=1, 2,3,4$)
to the Lagrangian wave Eq.(\ref{2}) and ignoring the higher terms,
one can obtain the following set of equations
\begin{eqnarray}
&DB&[c_{1}(\partial_{0}\tilde{I}_{o})(\partial_{1}\tilde{I}_{o})
-c_{1}\beta(\partial_{0}\tilde{I}_{o})^{3}(\partial_{1}\tilde{I}_{o})+
c_{1}\beta(\partial_{1}\tilde{I}_{o})(\partial_{0}\tilde{I}_{o})^{2}
eA_{0}\nonumber\\
&-&c_{0}\beta(\partial_{1}\tilde{I}_{o})^{4}
+e A_{0}c_{1}(\partial_{1}\tilde{I}_{o})-
c_{0}(\partial_{1}\tilde{I}_{o})^{2}]+BF[c_{3}(\partial_{1}\tilde{I}_{o})^{2}+
c_{3}\beta(\partial_{1}\tilde{I}_{o})^{3}
\nonumber\\&-&c_{1}(\partial_{1}
\tilde{I}_{o})(\partial_{3}\tilde{I}_{o})-c_{1}\beta(\partial_{3}\tilde{I}_{o})^{3}
(\partial_{1}\tilde{I}_{o})-eA_{3}c_{1}\beta(\partial_{3}\tilde{I}_{o})^{2}
(\partial_{1}\tilde{I}_{o})]+DC^{-1}
\nonumber\\
&\times&[c_{2}\beta eA_{0}(\partial_{0}\tilde{I}_{o})^{2}(\partial_{2}\tilde{I}_{o})
+c_{2}\beta(\partial_{0}\tilde{I}_{o})^{3}(\partial_{2}\tilde{I}_{o})-
c_{0}\beta(\partial_{2}\tilde{I}_{o})^{2}
-c_{0}\beta(\partial_{2}\tilde{I}_{o})^{4}
\nonumber\\
&+&c_{2}eA_{0}(\partial_{2}\tilde{I}_{o})+c_{2}(\partial_{0}\tilde{I}_{o})
(\partial_{2}\tilde{I}_{o})]
+F^{-1}[c_{3}(\partial_{2}\tilde{I}_{o})^{2}
+c_{3}\beta(\partial_{2}\tilde{I}_{o})^{4}\nonumber\\
&-&c_{2}(\partial_{3}\tilde{I}_{o})(\partial_{2}\tilde{I}_{o})-
c_{2}\beta(\partial_{2}\tilde{I}_{o})(\partial_{3}\tilde{I}_{o})^{3}
+c_{2}eA_{3}(\partial_{2}\tilde{I}_{o})+c_{2}eA_{3}\beta
(\partial_{3}\tilde{I}_{o})^{2}
\nonumber\\
&\times&(\partial_{2}\tilde{I}_{o})]+[c_{3}(\partial_{3}\tilde{I}_{o})
(\partial_{0}\tilde{I}_{o})-c_{3}\beta(\partial_{0}\tilde{I}_{o})^{3}(\partial_{3}\tilde{I}_{o})
-c_{0}(\partial_{3}\tilde{I}_{o})^{2}-c_{0}\beta(\partial_{3}
\tilde{I}_{o})^{4}\nonumber\\
&-&c_{0}eA_{3}(\partial_{3}\tilde{I}_{o})-c_{0}eA_{3}\beta(\partial_{3}
\tilde{I}_{o})^{3}+c_{3}eA_{0}(\partial_{3}
\tilde{I}_{o})+c_{3}eA_{0}\beta(\partial_{0}\tilde{I}_{o})^{2}
(\partial_{3}\tilde{I}_{o})]\nonumber\\
&+&eA_{3}[c_{3}(\partial_{0}\tilde{I}_{o})+c_{3}\beta(\partial_{0}\tilde{I}_{o})^{3}-c_{0}(\partial_{3}
\tilde{I}_{o})+c_{0}\beta(\partial_{3}\tilde{I}_{o})^{3}-c_{3}eA_{0}
\nonumber\\
&-&c_{3}\beta(\partial_{0}\tilde{I}_{o})^{2}+c_{0}eA_{3}+c_{0}eA_{3}
\beta(\partial_{3}\tilde{I}_{o})^{2}]-Dm^{2}c_{0}-Fm^{2}c_{3}=0,\label{6}\\
&D&[c_{1}(\partial_{0}\tilde{I}_{o})^{2}
+c_{1}\beta(\partial_{0}\tilde{I}_{o})^{4}-
c_{0}(\partial_{0}\tilde{I}_{o})(\partial_{1}\tilde{I}_{o})+c_{0}
\beta(\partial_{0}\tilde{I}_{o})(\partial_{1}\tilde{I}_{o})^{3}
\nonumber\\
&-&e A_{0}c_{1}(\partial_{0}\tilde{I}_{o})+c_{1}\beta(\partial_{0}\tilde{I}_{o})^{3}
eA_{0}]+F[c_{3}(\partial_{0}\tilde{I}_{o})(\partial_{1}\tilde{I}_{o})+
c_{3}\beta(\partial_{1}\tilde{I}_{o})^{3}(\partial_{0}\tilde{I}_{o})\nonumber\\
&-&c_{1}(\partial_{0}
\tilde{I}_{o})(\partial_{3}\tilde{I}_{o})
-c_{1}\beta(\partial_{3}\tilde{I}_{o})^{3}
(\partial_{0}\tilde{I}_{o})-c_{1}eA_{3}(\partial_{0}\tilde{I}_{o})
-eA_{3}c_{1}\beta(\partial_{3}\tilde{I}_{o})^{2}
(\partial_{0}\tilde{I}_{o})]\nonumber\\
&+&(ZD+F^{2})C^{-1}[
c_{2}(\partial_{1}\tilde{I}_{o})(\partial_{2}\tilde{I}_{o})
+c_{2}\beta(\partial_{1}\tilde{I}_{o})^{3}(\partial_{2}\tilde{I}_{o})-
c_{1}\beta(\partial_{2}\tilde{I}_{o})^{2}\nonumber\\
&-&c_{1}\beta(\partial_{2}\tilde{I}_{o})^{4}]
-m^{2}c_{1}+eA_{0}D[c_{1}(\partial_{0}\tilde{I}_{o})
-c_{1}\beta(\partial_{0}\tilde{I}_{o})^{3}
-c_{0}(\partial_{1}\tilde{I}_{o})
\nonumber\\
&-&c_{0}\beta(\partial_{1}\tilde{I}_{o})^{3}+c_{1}eA_{0}+c_{1}eA_{0}\beta
(\partial_{0}\tilde{I}_{o})^{2}]
+eA_{0}F[c_{3}(\partial_{1}\tilde{I}_{o})
+c_{3}\beta(\partial_{0}\tilde{I}_{o})^{3}\nonumber\\
&-&c_{1}(\partial_{3}\tilde{I}_{o})-c_{0}\beta(\partial_{3}
\tilde{I}_{o})^{3}-c_{1}eA_{3}-c_{1}eA_{3}\beta(\partial_{3}\tilde{I}_{o})^{2}]
+eA_{0}A[c_{3}(\partial_{1}
\tilde{I}_{o})\nonumber\\
&+&c_{3}\beta(\partial_{1}\tilde{I}_{o})^{3}-c_{1}(\partial_{3}
\tilde{I}_{o})-c_{1}\beta(\partial_{3}\tilde{I}_{o})^{3}-c_{1}eA_{3}
-c_{1}\beta(\partial_{3}\tilde{I}_{o})^{2}eA_{3}]\nonumber
\end{eqnarray}
\begin{eqnarray}
&+&eA_{3}F[c_{1}(\partial_{0}
\tilde{I}_{o})+
c_{1}\beta(\partial_{0}\tilde{I}_{o})^{3}-c_{0}(\partial_{1}
\tilde{I}_{o})-c_{0}\beta(\partial_{1}\tilde{I}_{o})^{3}+c_{1}eA_{0}\nonumber\\
&+&c_{1}\beta(\partial_{0}\tilde{I}_{o})^{2}eA_{0}]=0,\label{7}\\
&F&[c_{3}(\partial_{0}\tilde{I}_{o})(\partial_{2}\tilde{I}_{o})
+c_{3}\beta(\partial_{0}\tilde{I}_{o})(\partial_{2}\tilde{I}_{o})^{3}-
c_{2}(\partial_{0}\tilde{I}_{o})(\partial_{3}\tilde{I}_{o})-c_{2}
\beta(\partial_{0}\tilde{I}_{o})\nonumber\\
&\times&(\partial_{3}\tilde{I}_{o})^{3}-eA_{3}c_{2}(\partial_{0}\tilde{I}_{o})
-c_{2}\beta(\partial_{3}\tilde{I}_{o})^{2}(\partial_{0}\tilde{I}_{o})
eA_{0}]+D[c_{2}(\partial_{0}\tilde{I}_{o})^{2}\nonumber\\
&-&c_{0}(\partial_{0}\tilde{I}_{o})(\partial_{2}\tilde{I}_{o})+
c_{2}\beta(\partial_{0}\tilde{I}_{o})^{4}
-c_{0}\beta(\partial_{2}\tilde{I}_{o})^{3}
(\partial_{0}\tilde{I}_{o})+c_{2}eA_{0}(\partial_{0}\tilde{I}_{o})\nonumber\\
&+&eA_{3}c_{2}\beta(\partial_{2}\tilde{I}_{o})^{2}(\partial_{0}
\tilde{I}_{o})]-B[c_{2}(\partial_{1}\tilde{I}_{o})^{2}
+c_{2}\beta(\partial_{1}\tilde{I}_{o})^{4}-
c_{1}\beta(\partial_{1}\tilde{I}_{o})(\partial_{2}\tilde{I}_{o})\nonumber\\
&-&c_{1}\beta(\partial_{1}\tilde{I}_{o})(\partial_{2}\tilde{I}_{o})^{3}]
+B[c_{3}(\partial_{3}\tilde{I}_{o})(\partial_{2}\tilde{I}_{o})
+c_{3}\beta(\partial_{3}\tilde{I}_{o})(\partial_{2}\tilde{I}_{o})^{3}-
c_{2}\beta(\partial_{3}\tilde{I}_{o})^{2}\nonumber\\
&-&c_{2}\beta(\partial_{3}\tilde{I}_{o})^{4}-eA_{3}c_{2}(\partial_{3}\tilde{I}_{o})
-c_{2}\beta(\partial_{3}\tilde{I}_{o})^{3}eA_{3}]
-F[c_{2}(\partial_{0}\tilde{I}_{o})(\partial_{3}\tilde{I}_{o})\nonumber\\
&+&c_{2}\beta(\partial_{3}\tilde{I}_{o})(\partial_{0}\tilde{I}_{o})^{3}-
c_{0}\beta(\partial_{2}\tilde{I}_{o})(\partial_{3}\tilde{I}_{o})
-c_{0}\beta(\partial_{2}\tilde{I}_{o})^{3}(\partial_{3}\tilde{I}_{o})
+e A_{0}c_{2}(\partial_{3}\tilde{I}_{o})\nonumber\\
&+&c_{2}\beta(\partial_{0}
\tilde{I}_{o})^{2}(\partial_{3}\tilde{I}_{o})eA_{0}]-(ZB+F^{2})m^{2}c_{2}
+eA_{0}F[c_{3}(\partial_{2}\tilde{I}_{o})
+c_{3}\beta(\partial_{2}\tilde{I}_{o})^{3}\nonumber\\
&-&c_{2}(\partial_{3}\tilde{I}_{o})
-c_{2}\beta(\partial_{3}\tilde{I}_{o})^{3}-c_{2}eA_{3}-c_{2}eA_{3}\beta
(\partial_{3}\tilde{I}_{o})^{2}]
+eA_{0}D[c_{2}(\partial_{0}\tilde{I}_{o})\nonumber\\
&+&c_{2}\beta(\partial_{0}\tilde{I}_{o})^{3}
-c_{0}(\partial_{2}\tilde{I}_{o})-c_{0}\beta(\partial_{2}
\tilde{I}_{o})^{3}+c_{2}eA_{0}+c_{2}eA_{0}\beta(\partial_{0}\tilde{I}_{o})^{2}]\nonumber\\
&+&eA_{3}B[c_{3}(\partial_{2}\tilde{I}_{o})+c_{3}\beta(\partial_{2}\tilde{I}_{o})^{3}-c_{2}(\partial_{3}
\tilde{I}_{o})-c_{2}\beta(\partial_{3}\tilde{I}_{o})^{3}
-c_{2}eA_{3}\nonumber\\
&-&c_{2}\beta eA_{3}(\partial_{3}\tilde{I}_{o})^{2}]
+eA_{3}F[c_{2}(\partial_{0}\tilde{I}_{o})+c_{2}\beta(\partial_{0}\tilde{I}_{o})^{3}
-c_{0}(\partial_{2}\tilde{I}_{o})-c_{0}\beta(\partial_{2}\tilde{I}_{o})^{3}\nonumber\\
&+&c_{2}eA_{0}+c_{2}\beta(\partial_{0}\tilde{I}_{o})^{2}eA_{0}]=0,\label{8}\\
&F&[c_{2}(\partial_{0}\tilde{I}_{o})(\partial_{3}\tilde{I}_{o})
+c_{2}\beta(\partial_{3}\tilde{I}_{o})(\partial_{0}\tilde{I}_{o})^{3}-
c_{0}\beta(\partial_{3}\tilde{I}_{o})(\partial_{2}\tilde{I}_{o})
-c_{0}\beta(\partial_{2}\tilde{I}_{o})^{3}\nonumber\\
&\times&(\partial_{3}\tilde{I}_{o})+eA_{0}c_{2}(\partial_{3}\tilde{I}_{o})
+c_{2}\beta(\partial_{0}\tilde{I}_{o})^{2}(\partial_{3}\tilde{I}_{o})eA_{0}]
[c_{3}(\partial_{0}\tilde{I}_{o})^{2}
+c_{3}\beta(\partial_{0}\tilde{I}_{o})^{4}\nonumber\\
&-&c_{0}(\partial_{0}\tilde{I}_{o})(\partial_{3}\tilde{I}_{o})-c_{0}\beta
(\partial_{0}\tilde{I}_{o})(\partial_{3}\tilde{I}_{o})^{3}
-e A_{3}c_{3}(\partial_{0}\tilde{I}_{o})+c_{3}\beta(\partial_{0}\tilde{I}_{o})^{3}
eA_{3}\nonumber\\
&-&c_{0}(\partial_{0}\tilde{I}_{o})eA_{3}-c_{0}\beta
(\partial_{3}\tilde{I}_{o})^{2}(\partial_{0}\tilde{I}_{o})eA_{3}]
+BF[c_{1}(\partial_{0}\tilde{I}_{o})(\partial_{1}\tilde{I}_{o})+c_{1}eA_{0}\nonumber\\
&\times&(\partial_{1}\tilde{I}_{o})-c_{0}(\partial_{1}
\tilde{I}_{o})^{2}-c_{0}\beta(\partial_{2}\tilde{I}_{o})^{4}
+c_{1}\beta(\partial_{0}\tilde{I}_{o})^{3}(\partial_{1}\tilde{I}_{o})+eA_{0}c_{1}
\beta(\partial_{0}\tilde{I}_{o})^{2}\nonumber\\
&\times&(\partial_{1}\tilde{I}_{o})]-ZB[c_{3}(\partial_{1}\tilde{I}_{o})^{2}
+c_{3}\beta(\partial_{1}\tilde{I}_{o})^{4}-c_{1}\beta(\partial_{1}
\tilde{I}_{o})(\partial_{3}\tilde{I}_{o})
-c_{1}\beta(\partial_{1}\tilde{I}_{o})\nonumber\\
&\times&(\partial_{3}\tilde{I}_{o})^{2}-c_{1}e A_{3}(\partial_{1}\tilde{I}_{o})
+c_{1}\beta e A_{3}(\partial_{3}\tilde{I}_{o})^{2}(\partial_{1}\tilde{I}_{o})]-B[c_{3}
(\partial_{2}\tilde{I}_{o})^{2}\nonumber\\
&-&c_{2}(\partial_{3}\tilde{I}_{o})(\partial_{2}\tilde{I}_{o})+c_{3}
\beta(\partial_{2}\tilde{I}_{o})^{4}-c_{2}\beta(\partial_{3}
\tilde{I}_{o})^{3}(\partial_{2}\tilde{I}_{o})-e A_{3}c_{2}(\partial_{2}\tilde{I}_{o})\nonumber\\
&+&c_{2}\beta(\partial_{3}\tilde{I}_{o})^{2}(\partial_{2}\tilde{I}_{o})eA_{3}]+
eA_{0}[c_{3}(\partial_{0}\tilde{I}_{o})
+c_{3}\beta(\partial_{0}\tilde{I}_{o})^{3}-c_{0}eA_{3}+c_{3}eA_{0}\nonumber\\
&-&c_{0}(\partial_{3}\tilde{I}_{o})-c_{0}\beta(\partial_{3}\tilde{I}_{o})^{3}
+c_{3}eA_{0}\beta
(\partial_{0}\tilde{I}_{o})^{2}+c_{0}eA_{3}\beta
(\partial_{3}\tilde{I}_{o})^{2}]\nonumber\\
&-&Fm^{2}c_{0}-Zm^{2}c_{3}=0.\label{9}
\end{eqnarray}
Using the variables separation technique, the particle's action is defined as follows
\begin{equation}
\tilde{I}_{o}=-(E-J\Omega_{H})t+R(r)+N\chi+\Theta(\theta),
\end{equation}
where $J$ and $E$ denotes the angular momentum and particles energy, respectively and
from the above Eqs.(\ref{6})-(\ref{9}), we get a
non-trivial matrix equation, i.e.,
\begin{equation}
V(c_{0},c_{1},c_{2},c_{3})^{t}=0,\label{121212}
\end{equation}
where $V$ is a $4\times4$ matrix, whose elements are given as below
\begin{eqnarray}
V_{00}&=&-\dot{R}^{2}BD-BD\beta(E-J\Omega_{H})^{4}-DC^{-1}N^{2}
+DC^{-1}\beta N^{4}-\dot{\Theta}^{2}\nonumber\\
&-&\beta\dot{\Theta}^{2}-eA_{3}\dot{\Theta}-\beta eA_{3}\dot{\Theta}^{3}
+m^{2}D-eA_{3}\dot{\Theta}+e^{2}A_{3}^{2}\beta\dot{\Theta}^{2}+e^{2}A_{3}^{2}\beta\dot{\Theta}^{3},\nonumber\\
V_{01}&=&DB[(E-J\Omega_{H})\dot{R}-\dot{R}\beta(E-J\Omega_{H})^{3}
+\dot{R}eA_{0}-\dot{R}\beta(E-J\Omega_{H})^{2}eA_{0}]\nonumber\\
&+&BF[\dot{R}\dot{\Theta}+\beta \dot{R}\dot{T}^{3}-eA_{3}\dot{R}+eA_{3}
\dot{\Theta}\dot{R}],\nonumber\\
V_{02}&=&DC^{-1}[(E-J\Omega_{H})N+\beta(E-J\Omega_{H})^{3}N-
eA_{0}N+\beta N eA_{0}\nonumber\\
&\times&(E-J\Omega_{H})^{2}]+FC^{-1}[N\dot{T}-\beta\dot{\Theta}^{3}N
-eA_{3}N+\beta eA_{3}\dot{\Theta}^{2}N],\nonumber\\
V_{03}&=&\beta eA_{0}N^{2}\dot{\Theta}-FC^{-1}[N^{2}-\beta \dot{\Theta}^{4}]
+(E-J\Omega_{H})\dot{\Theta}-\beta(E-J\Omega_{H})^{3}\dot{\Theta}\nonumber\\
&-&eA_{0}\dot{\Theta}+eA_{3}[(E-J\Omega_{H})-\beta(E-J\Omega_{H})^{3}
+eA_{0}+\beta(E-J\Omega_{H})^{2}]\nonumber\\
&-&BD[{\dot R}^{2}-\beta{\dot R}^{3}]-m^{2}F,\nonumber\\
V_{10}&=&D[(E-J\Omega_{H})\dot{R}+\beta(E-J\Omega_{H})\dot{R}^{3}]+
eA_{0}D[-\dot{R}+\beta\dot{R}^{3}]\nonumber\\
&+&eA_{3}F[-\dot{R}+\beta \dot{R}^{3}],\nonumber\\
V_{11}&=&D[(E-J\Omega_{H})^{2}-\beta(E-J\Omega_{H})^{4}+(E-J\Omega_{H})eA_{0}+
\beta eA_{0}\nonumber\\
&\times&(E-J\Omega_{H})^{3}]
+F[(E-J\Omega_{H})\dot{\Theta}-\beta(E-J\Omega_{H})\dot{\Theta}^{3}+(E-J\Omega_{H})eA_{3}\nonumber\\
&-&\beta\dot{\Theta}^{2}(E-J\Omega_{H})eA_{3}]
+(ZD+F^{2})C^{-1}[-N^{2}+\beta N^{4}]-m^{2}\nonumber\\
&+&eA_{0}D[-(E-J\Omega_{H})+\beta(E-J\Omega_{H})^{3}+eA_{0}-eA_{0}\beta(E-J\Omega_{H})^{2}]\nonumber\\
&+&eA_{0}F[-\dot{\Theta}+\beta\dot{\Theta}^{3}-eA_{3}+\beta\dot{\Theta}^{2}eA_{3}]
+eA_{3}A[-\dot{\Theta}+\beta \dot{\Theta}^{3}-eA_{3}\nonumber\\
&-&\beta\dot{\Theta}^{2}eA_{3}]+eA_{3}F[-(E-J\Omega_{H})+\beta(E-J\Omega_{H})^{3}+eA_{0}\nonumber\\
&-&\beta (E-J\Omega_{H})eA_{0}],\nonumber\\
V_{12}&=&(ZD+F^{2})C^{-1}[\dot{R}N-\beta\dot{R}N],\nonumber\\
V_{13}&=&F[-(E-J\Omega_{H})\dot{R}+\beta(E-J\Omega_{H})\dot{R}^{3}]-eA_{0}F[\dot{R}-\beta\dot{R}^{3}]\nonumber\\
&+&eA_{3}Z[\dot{R}-\beta\dot{R}^{3}],\nonumber\\
V_{20}&=&D[-(E-J\Omega_{H})N-\beta(E-J\Omega_{H}) N^{3}]-F[-N\dot{\Theta}+\beta \dot{\Theta}N^{3}]\nonumber\\
&+&eA_{0}D[N-\beta N^{3}]+eA_{3}[-N+\beta N^{3}],\nonumber\\
V_{21}&=&B[\dot{R}N-\beta\dot{R}N^{3}],\nonumber\\
V_{22}&=&F[(E-J\Omega_{H})\dot{\Theta}-(E-\Omega_{H}J)\beta
\dot{\Theta}^{3}+eA_{0}(E-J\Omega_{H})-\beta eA_{3}\nonumber\\
&\times&(E-J\Omega_{H})\dot{\Theta}^{2}]+D[(E-J\Omega_{H})^{2}-
\beta(E-J\Omega_{H})^{4}-eA_{0}(E-J\Omega_{H})\nonumber\\
&+&\beta(E-J\Omega_{H})
N^{2}]-B[\dot{R}^{2}+\beta \dot{R}^{4}]+B[\dot{\Theta}^{2}+\beta\dot{\Theta}^{4}-eA_{3}\dot{\Theta}+
\beta eA_{3}\dot{S}^{3}]\nonumber\\
&+&F[(E-J\Omega_{H})
\dot{\Theta}-\beta(E-J\Omega_{H})^{3}\dot{\Theta}-eA_{0}\dot{\Theta}+\beta(E-J\Omega_{H})^{2}\dot{\Theta}eA_{0}]\nonumber
\end{eqnarray}
\begin{eqnarray}
&-&(ZD+F^{2})m^{2}+F eA_{0}[-\dot{\Theta}+\beta\dot{\Theta}^{3}-eA_{3}-\beta\dot{\Theta}^{2}eA_{3}]\nonumber\\
&+&DeA_{0}[-(E-J\Omega_{H})+\beta(E-J\Omega_{H})^{3}+eA_{0}+\beta eA_{0}(E-J\Omega_{H})^{2}]\nonumber\\
&+&B eA_{3}[-\dot{S}+\beta\dot{S}^{3}-eA_{3}-\beta\dot{S}^{2}eA_{3}]
+eA_{3}[-(E-J\Omega_{H})\nonumber\\
&+&\beta(E-J\Omega_{H})^{3}+eA_{0}+\beta eA_{0}(E-J\Omega_{H})^{2}],\nonumber\\
V_{23}&=&F[(E-J\Omega_{H})N+\beta(E-J\Omega_{H}) N^{3}]+[\dot{\Theta}N-\beta N^{3}\dot{\Theta}]B\nonumber\\
&-&eA_{0}F[\beta N^{3}-N]+eA_{3}B[\beta N^{3}-N],\nonumber\\
V_{30}&=&[-\dot{\Theta}-\beta(E-J\Omega_{H})
\dot{\Theta}^{3}+eA_{3}(E-J\Omega_{H})-\beta eA_{3}(E-J\Omega_{H})\dot{\Theta}^{2}]\nonumber\\
&-&BF[\dot{R}^{2}-\beta(E-J\Omega_{H})^{3}\dot{R}+\beta\dot{R}^{4}]+F[-\dot{T}N
+\beta N^{4}]+m^{2}F\nonumber\\
&+&eA_{0}[-\dot{\Theta}+\beta \dot{\Theta}^{3}-e
A_{3}-\beta eA_{3}\dot{\Theta}^{2}],\nonumber\\
V_{31}&=&BF[-(E-J\Omega_{H})\dot{R}+\beta(E-J\Omega_{H})^{3}
\dot{R}+eA_{0}\dot{\Theta}-\beta eA_{0}(E-J\Omega_{H})^{2}\dot{\Theta}]\nonumber\\
&-&ZB[-\dot{R}\dot{\Theta}+\beta\dot{R}\dot{\Theta}^{2}-\dot{\Theta}eA_{3}-
\beta\dot{\Theta}^{2}\dot{R}eA_{3}],\nonumber\\
V_{32}&=&F[-(E-J\Omega_{H})\dot{\Theta}+\beta(E-J\Omega_{H})^{3}
\dot{\Theta}+eA_{0}\dot{\Theta}-\beta \dot{\Theta}eA_{0}(E-J\Omega_{H})^{2}]\nonumber\\
&+&B[N\dot{\Theta}-\beta N\dot{\Theta}^{3}
-NeA_{3}+\dot{\Theta}^{2}\beta NeA_{3}],\nonumber\\
V_{33}&=&[(E-J\Omega_{H})^{2}-\beta(E-J\Omega_{H})^{4}+eA_{3}+\beta
(E-J\Omega_{H})^{3}eA_{3}]\nonumber\\
&-&ZB[\dot{R}^{2}-\beta\dot{R}^{4}]-B[N^{2}
+\beta N^{4}]-m^{2}Z+eA_{0}[-(E-J\Omega_{H})\nonumber\\
&+&\beta(E-J\Omega_{H})^{3}+e
A_{0}-\beta eA_{0}(E-J\Omega_{H})^{2}],\nonumber
\end{eqnarray}
where $\dot{R}=\partial_{r}\tilde{I}_{o}$,
$\dot{\Theta}=\partial_{\theta}\tilde{I}_{o}$ and $N=\partial_{\chi}\tilde{I}_{o}$.
For the non-trivial solution, we put $\det(\bf{V})=0$ and computing the radial
part for resultant components, the following integral can be obtained as follows
\begin{equation}\label{a1}
R^{\pm}=\pm \int\sqrt{\frac{(E-J{\Omega_{H}}-eA_{0})^{2}
+X_{2}(1+\frac{X_{1}}{X_{2}}\beta)}{B}}dr,
\end{equation}
where $R^-$ and $R^+$ denote the radial function of
incoming and outgoing boson particles, respectively.
Although the function $X_{1}$ and $X_{2}$ can be
determined as
\begin{eqnarray}
X_{1}&=&eA_{3}\frac{F}{D}[(E-J\Omega_{H})^{3}-(E-J\Omega_{H})eA_{0}]
+(ZD+F^{2})(CD)^{-1}N^{4}\nonumber\\&+&
\beta(E-J\Omega_{H})^{3}eA_{0}+eA_{0}[(E-J\Omega_{H})^{3}-eA_{0}(E
-J\Omega_{H})^{2}]
\nonumber\\&-&\frac{F}{D}\dot{\Theta}^{2}(E-J\Omega_{H})eA_{3}
+\frac{F}{D}(E-J\Omega_{H})\dot{\Theta}^{3}-(E-J\Omega_{H})^{4}
+eA_{0}\frac{F}{D}\nonumber\\
&\times&[\dot{\Theta}^{3}+\dot{\Theta}^{2}eA_{3}]
+eA_{3}Z[\dot{\Theta}^{3}-\dot{\Theta}^{2}eA_{3}],\nonumber
\end{eqnarray}
and
\begin{eqnarray}
X_{2}&=&\frac{F}{D}[(E-J\Omega_{H})\dot{\Theta}+(E-J\Omega_{H})eA_{3}]
-(ZD+F^{2})(CD)^{-1}N^{2}-m^{2}
\nonumber\\
&+&eA_{0}\frac{F}{D}[-\dot{\Theta}-eA_{3}]
-eA_{3}\frac{Z}{D}\dot{\Theta}-eA_{3}
+eA_{3}\frac{F}{D}[-(E-J\Omega_{H})+eA_{0}].\nonumber
\end{eqnarray}
After applying taylor's series the functions $Z(r)$ and $B(r)$ near the horizon can be obtained as
\begin{equation}\label{a2}
Z(r_+)\approx (r-r_+)Z'(r_+),\quad\quad B(r_+)\approx (r-r_+)B'(r_+).
\end{equation}
By using above relations in Eq.(\ref{a1}), we consider that the leading wave equation
has two poles at $r=r_+$.
Utilizing Eqs.(\ref{a1}) and (\ref{a2}), by integrating around the pole, we get
\begin{equation}
ImR^{\pm}
=\pm i\pi\frac{E-{\Omega_{H}}J-eA_{0}}{2\kappa(r_{+})(1+\Xi\beta)},
\end{equation}
where
$\Xi=6 \left(m^{2}+\frac{\left(J^{2}_{\theta}
+J^{2}_{\phi}\csc^{2}\theta\right)}{r_{+}^{2}}
\right)>0$.

The surface gravity $\kappa(r_{+})$ is given as follows \cite{R6}
\begin{eqnarray}
\kappa(r_{+})&=&\left[\frac{\frac{l\alpha
}{\omega}(\tilde{g}^{2}+\tilde{e}^{2}+\omega^{2}\tilde{k})
+\frac{\omega^{2}\tilde{k}}{a^{2}-l^{2}}r_{+}-M}{r^{2}_{+}+(l+a)^{2}}
\times\left(\frac{(a-l)\alpha}{\omega}r_{+}+1\right)\right.\nonumber\\
&\times&\left.\left(1-\frac{(a+l)\alpha}{\omega}r_{+}\right)\right].
\nonumber
\end{eqnarray}
The corrected tunneling probability $(\Gamma)$ for boson particles can be obtained as
\begin{eqnarray}\nonumber
\Gamma&=&\frac{Prob{[emission]}}{Prob{[absorption]}}=
\frac{\textmd{exp}[-2(ImR^++Im\Theta)]}{\textmd{exp}[-2(ImR^{-}-Im\Theta)]}
={\textmd{exp}[-4ImR^+]},\\\nonumber
&=&\exp\left[
\frac{-2\pi(E-J{\Omega_{H}}-eA_{0})(1+\beta\Xi)}{\frac{\frac{\alpha
l}{\omega}(\tilde{g}^{2}+\tilde{e}^{2}+\omega^{2}\tilde{k})
-M+\frac{\omega^{2}\tilde{k}}{a^{2}-l^{2}}r_{+}}{r^{2}_{+}+(l+a)^{2}}
\times\left(1-\frac{\alpha(l-a)}{\omega}r_{+}\right)\times
\left(1-\frac{\alpha(l+a)}{\omega}r_{+}\right)}\right].
\end{eqnarray}
We calculate the corrected Hawking temperature by
comparing the tunneling probability with Boltzmann factor, i.e., $\Gamma_B= e^{-(E-J\Omega_{H}-e A_0)/T'_H}$
\begin{eqnarray}
T^{\prime}_{H}&=&\left[\frac{{\Big(\frac{l\alpha
}{\omega}(\tilde{g}^{2}+\tilde{e}^{2}+\omega^{2}\tilde{k})
+\frac{\omega^{2}\tilde{k}}{a^{2}-l^{2}}r_{+}-M\Big)}
\Big(1-\frac{(l-a)\alpha}{\omega}r_{+}\Big)\Big(1-\frac{(l+a)\alpha}{\omega}r_{+}\Big)
}{2\pi\Big(r^{2}_{+}+(l+a)^{2}\Big)(1+\Xi\beta)}\right],\nonumber
\end{eqnarray}
\begin{eqnarray}
T^{\prime}_{H}&=&\left[\frac{{\Big(\frac{l\alpha
}{\omega}(\tilde{g}^{2}+\tilde{e}^{2}+\omega^{2}\tilde{k})
+\frac{\omega^{2}\tilde{k}}{a^{2}-l^{2}}r_{+}-M\Big)}
\Big(1-\frac{(l-a)\alpha}{\omega}r_{+}\Big)\Big(1-\frac{(l+a)\alpha}{\omega}r_{+}\Big)
}{2\pi\Big(r^{2}_{+}+(l+a)^{2}\Big)}\right]\nonumber\\
&\times&\left[1-\beta\Xi+(\beta\Xi)^2+...\right], \nonumber
\end{eqnarray}
by considering only first order quantum corrections, we can write
\begin{equation}
T^{\prime}_{H}=T_H\left[1-\beta\Xi\right],\label{aTH}
\end{equation}
where the semi-classical Hawking temperature $T_H$ is given as follows
\begin{equation}
T_{H}=\left[\frac{{\Big(\frac{l\alpha
}{\omega}(\tilde{g}^{2}+\tilde{e}^{2}+\omega^{2}\tilde{k})
+\frac{\omega^{2}\tilde{k}}{a^{2}-l^{2}}r_{+}-M\Big)}
\Big(1-\frac{(l-a)\alpha}{\omega}r_{+}\Big)\Big(1-\frac{(l+a)\alpha}{\omega}r_{+}\Big)
}{2\pi\Big(r^{2}_{+}+(l+a)^{2}\Big)}\right].\nonumber
\end{equation}
The corrected tunneling probability depends on $A_{0}$, $E$, $\ell$,
$\omega$, $e$, $\alpha$, $J$ and $\beta$, i.e., vector potential of BHs,
energy of particle, $NUT$ parameter, kerr-like rotation parameter,
charges of particles, acceleration of BHs, angular momentum of
particle and correction parameter, respectively. We can observe
that the corrected Hawking temperature does not only
depend upon the BH properties but also depends upon the mass and angular momentum
of the radiated particles and quantum corrections $\beta$.
It is to be noted that the first order correction term is same as
semi-classical original Hawking term $T_H$, while the
next order correction term must be smaller than the preceding
term satisfying GUP.

It is to be noted that the corrected temperature of boson
particles given in Eq.(\ref{aTH}) reduced to the temperature
of fermion particles in Eq.(4.20) for $(\beta=0)$ of Ref.\cite{R6}. So, the corrected
temperature depends upon the correction parameter.
Also, for $\ell=0$ and $\tilde{k}=1$, the above result reduces to the Hawking
temperature of accelerating and rotating BHs
with electric and magnetic charges \cite{[12]}. For $\alpha=0$, we
recover the temperature for non-accelerating BHs \cite{[13]}.
Moreover, for $\beta=0$, $\ell=0$, $\tilde{k}=1$ \& $\alpha=0$
in Eq.(\ref{aTH}), the Hawking temperature of the Kerr-Newman BH \cite{q6} is
recovered and which is reduced for $a=0$ to the temperature of ReissnerNordstr$\ddot{o}$m BH,
for $Q=0$, the temperature exactly reduces to
the Hawking temperature of the Schwarzschild BH \cite{[25]}.
In order to calculate the residual mass of BH, the temperature can be expressed as
\begin{equation}
T'_{H}=\frac{1}{8\pi M}\left[
1-6\beta\left(m^{2}+\frac{\left(J^{2}_{\theta}
+J^{2}_{\phi}\csc^{2}\theta\right)}{r_{+}^{2}}
\right)\right],\nonumber
\end{equation}
where $\left(m^{2}+\frac{\left(J^{2}_{\theta}
+J^{2}_{\phi}\csc^{2}\theta\right)}{r_{+}^{2}}
\right)$ denotes the kinetic energy component of radiated particles
related with tangent plane at horizon. For residual mass, we approximate the kinetic energy component as $\omega^2$.
Quantum corrections decelerate the increase in temperature during the radiation process.
These corrections cause the radiation ceased at some specific temperature, leaving the remnant mass. The temperature stops increasing when this condition holds \cite{Chen:2013pra}
\begin{equation}
(M-dM)(1+\beta\Xi)\simeq M.\nonumber
\end{equation}
For $dM=\omega$, $\beta=\frac{\beta_0}{M_{p}^2}$ and
$\omega\simeq M_{p}$ where $M_{p}$ is the Planck mass and
$\beta_0$ is a dimensionless parameter representing quantum gravity effects and $\beta_0<10^{5}$ \cite{Chen:2013tha,Liberati:2012jf}, one can obtain the following constraints
\begin{equation}
M_{Res} \simeq \frac{M_{p}^2}{\beta_0\omega}\gtrsim \frac{M_{p}}{\beta_0},
~~~~~~~~~~~T_{Res}\lesssim\frac{\beta_0}{8\pi M_p}\nonumber.
\end{equation}
It is important to mention here that the value of the corrected Hawking temperature is smaller than
the original temperature and BH stops radiating, when the mass of the BH reaches to its minimal
value $M_{Res}$.

\subsection{Graphical Analysis of $T'_{H}$ versus $r_{+}$}

This subsection is devoted to study the graphical behavior of Hawking temperature $T'_{H}$
with respect to horizon $r_{+}$ under quantum gravity effects. Furthermore, we study the physical
significance of these graphs under the influence of correction parameter $\beta$,
$NUT$ parameter $\ell$, rotation parameters $a$ and $\omega$, BH acceleration $\alpha$,
electric and magnetic charges $e$ and $g$, arbitrary parameter $k$,
for fixed BH mass $M=1$ and arbitrary parameter $\Xi=0.01$. We analyze stability and instability of
accelerating and rotating BH associated with $NUT$ parameter.
\begin{figure}\begin{center}
\epsfig{file=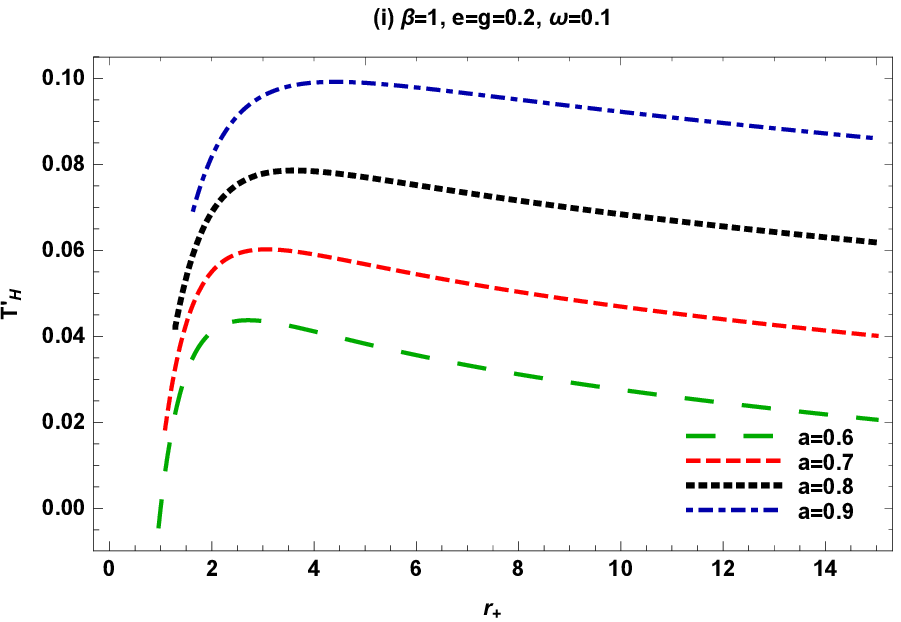,width=0.50\linewidth}\epsfig{file=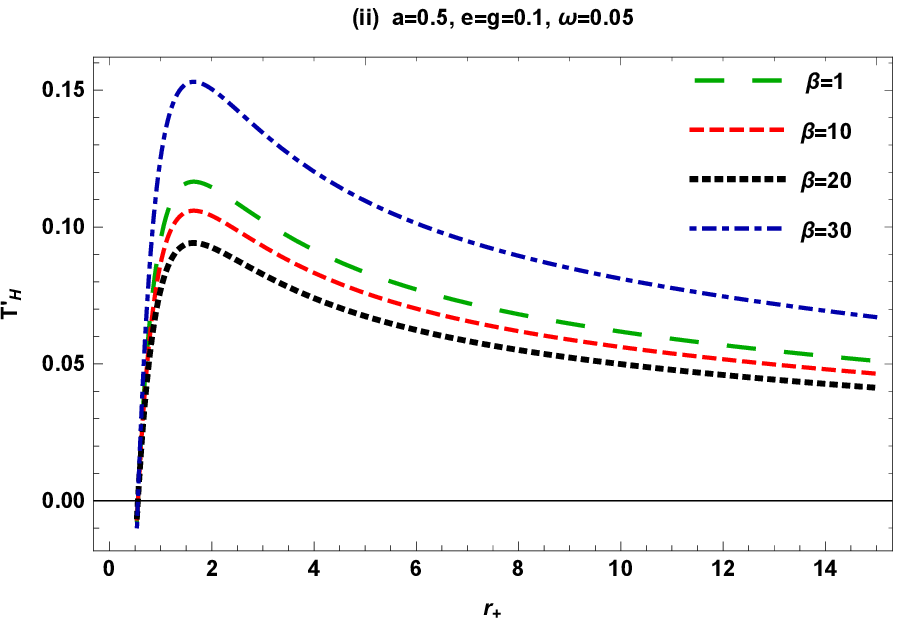,width=0.50\linewidth}
\caption{$T'_{H}$ vs $r_{+}$ for $\alpha=0.1$, $k=0.5$ \& $\ell=0.4$.}
\epsfig{file=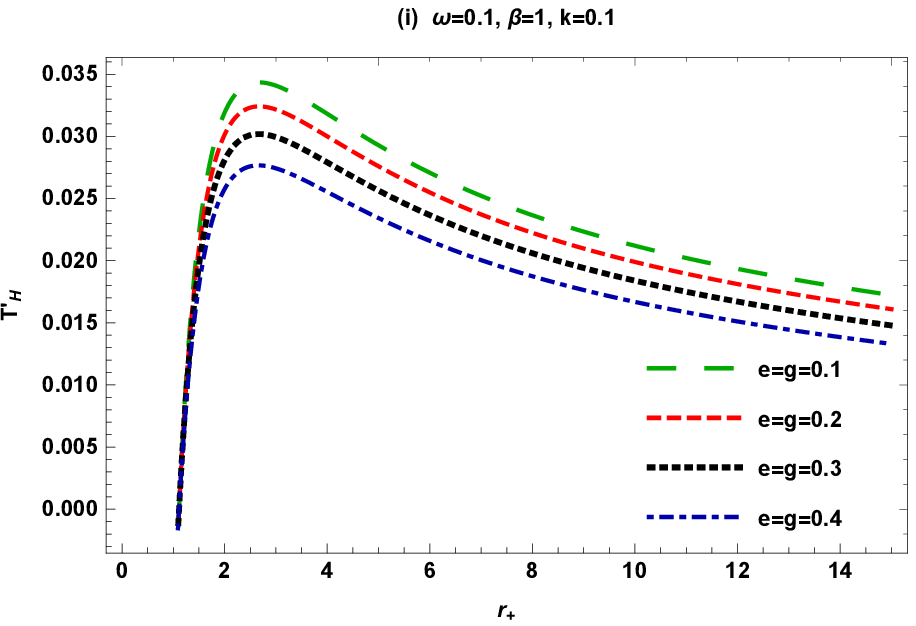,width=0.50\linewidth}\epsfig{file=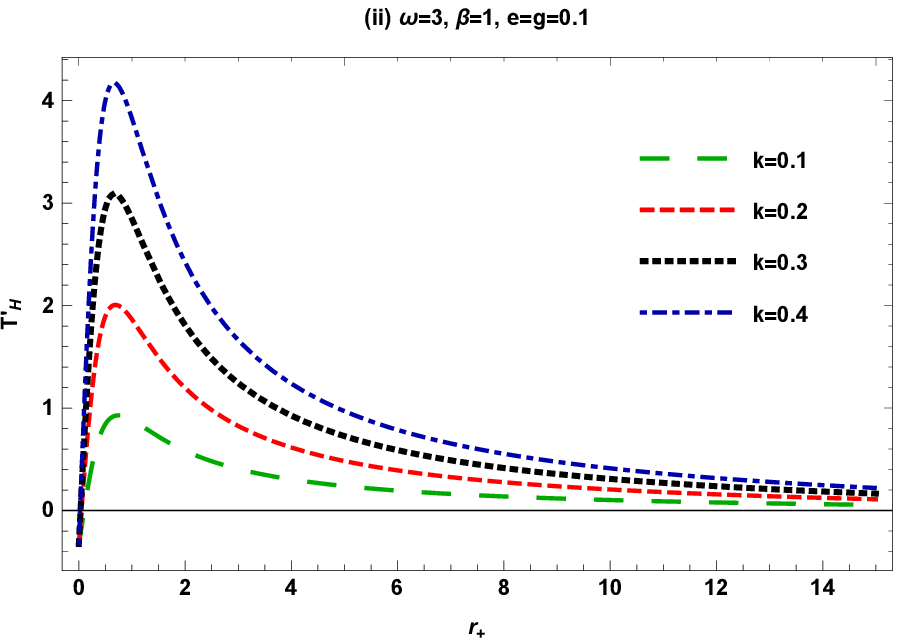,width=0.50\linewidth}
\caption{$T'_{H}$ vs $r_{+}$ for $\alpha=0.1$, $a=0.5$, \& $\ell=0.4$.}
\epsfig{file=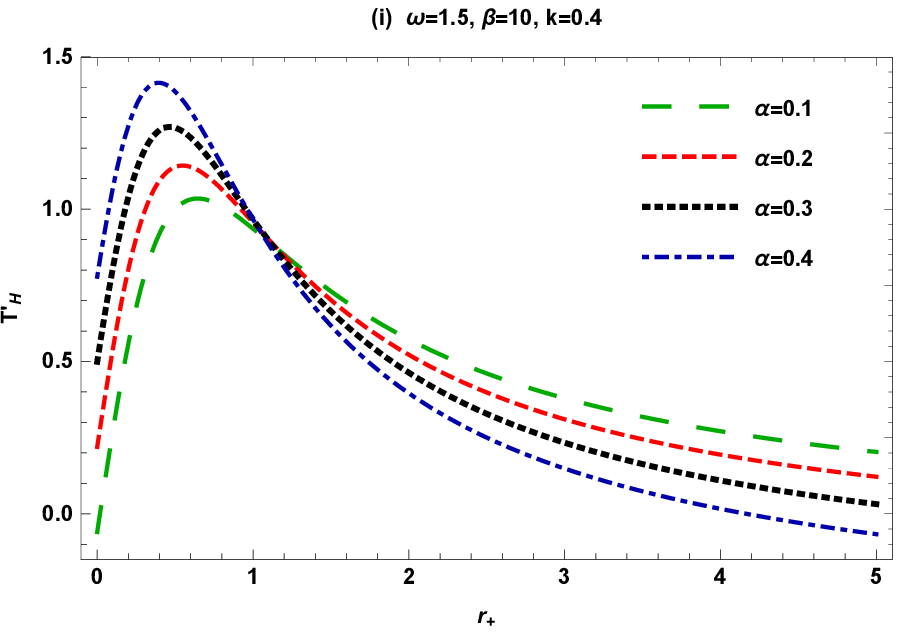,width=0.50\linewidth}\epsfig{file=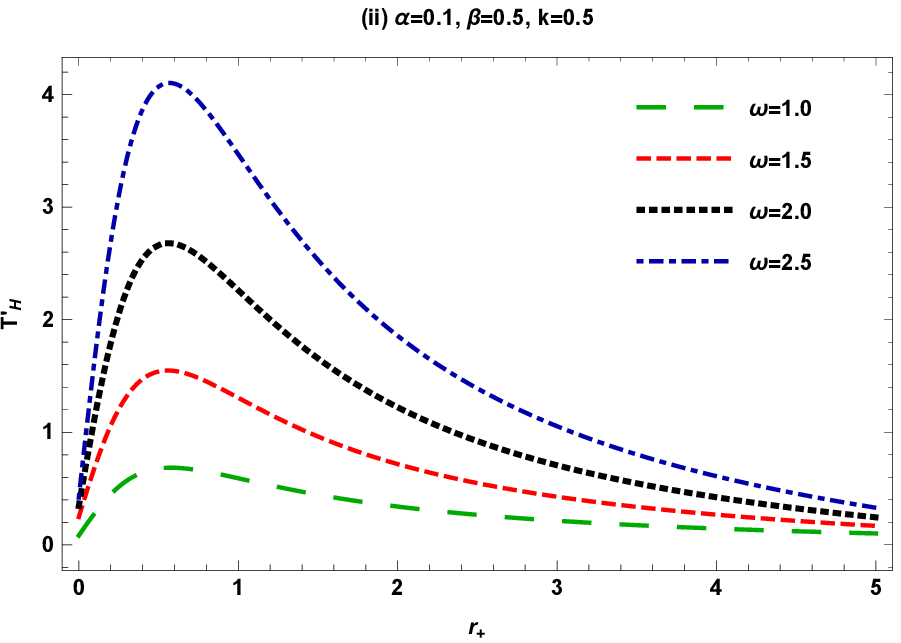,width=0.50\linewidth}
\caption{$T'_{H}$ vs $r_{+}$ for $a=0.5$, $e=g=5$ \& $\ell=0.4$.}
\end{center}
\end{figure}

\textbf{Figure 1} indicates the behavior of $T'_{H}$ w.r.t $r_{+}$
for fixed $\alpha=0.1$, $k=0.5$ and $\ell=0.4$ \cite{Chen:2013tha,Liberati:2012jf}.

\textbf{(i)} There can be seen an exponential
increase in the temperature $T'_{H}$
and after attaining a height the $T'_{H}$ slightly decreases
with the increasing horizon $r_{+}$ for varying values of rotation parameter $a$.
The $T'_{H}$ decreases as horizon increases, this physical
behavior reflects the BH stability with positive temperature till
$r_{+}\rightarrow\infty$. The temperature increases with the increase of $a$.

\textbf{(ii)} We observe the behavior of temperature for
varying values of correction parameter $\beta$ with fixed values of different parameters.
The $T'_{H}$ decreases with increasing horizon in the domain $0\leq r_{+}\leq15$
after attaining a maximum height. The maximum temperature with non-zero horizon reflects the BH remnant.
This physical behavior of $T'_{H}$ indicates the BH stability with positive range.
It is also worthy to note that the $T'_{H}$ increases with the
increase of correction parameter $\beta$.

\textbf{Figure 2} shows the behavior of $T'_{H}$ w.r.t $r_{+}$ for
fixed $\alpha=0.1$, $a=0.5$ \& $\ell=0.4$.

\textbf{(i)} Figure shows that the $T'_{H}$ exponentially increases and slowly falls down from a height
for different values of electric and magnetic charges. The decrease in $T'_{H}$ with
the increasing horizon exhibits the stable state of BH in the domain $0\leq r_{+}\leq15$.
It can be seen that the $T'_{H}$ decreases with the increase of BH electric
and magnetic charges $e$ and $g$, respectively.

\textbf{(ii)} Figure indicates that the temperature eventually
drops down from a height for different values of arbitrary parameter $k$.
There is a significant change in temperature as it decreases
exponentially and attains an asymptotically flat shape which indicates BH
stability till $r_{+}\rightarrow+\infty$. The $T'_{H}$ increases with increase of $k$.

\textbf{Figure 3} depicts the behavior of $T'_{H}$ w.r.t $r_{+}$ for
fixed $a=0.5$, $e=g=5$ \& $\ell=0.4$ for varying $a$ and $\omega$ in the domain $0\leq r_{+}\leq5$.

\textbf{(i)} We can observe that initially the temperature increases with increasing horizon
and after a maximum height it exponentially decreases with increasing which indicates the stable state
of BH for different values of BH acceleration $\alpha$.

\textbf{(ii)} There can be seen that $T'_{H}$ exponentially increases
and eventually falls down from a height and decreases as horizon increases
till $r_{+}\rightarrow+\infty$. This physical behavior identify the stability of
BH with positive range for different values of $\omega$. We can observe that with the
increase in the value of $\omega$ the $T'_{H}$ increases.

\section{Conclusion and Discussion}

In this paper, we have studied the quantum gravity effects for spin-1
(boson) particles from charged accelerating rotating BH having $NUT$ parameter. For this purpose, by considering
the GUP effects, we have used modified Lagrangian equation incorporating quantum effects
described the motion of spin-1 particles. Later, by applying the Hamilton-Jacobi
technique, we have calculated the tunneling probabilities of boson particles.
Moreover, we have analyzed the corrected Hawking temperatures
of these BHs. We have concluded that the modified tunneling probabilities are not just depend
upon the BHs properties but also depend on the properties of emitted boson
particles, i.e., energy, potential, surface gravity, particles charge and total angular
momentum. Moreover, it is important to note that the modified
tunneling probabilities as well as Hawking temperature depend on the quantum
particles which contributes gravitational radiation in form of massive particles (BH's energy carrier) tunneling.

When the quantum gravity effects are neglected, i.e., $(\beta=0)$, then the corrected
Hawking temperature Eq.(\ref{aTH}) is reduced to the
absolute temperature obtained by quantum tunneling of boson (spin-1) particles
\cite{A1}. If we ignore the potential effects $(A=0)$, the
modified Hawking temperature reduced to the temperature
of vector (spin-1) particles provided in Refs.\cite{R5,R30}.
Also, for $\ell=0$ and $\tilde{k}=1$, the above results are reduced to the Hawking
temperature of the accelerating and rotating BHs
with electric and magnetic charges \cite{[12]}. For $\alpha=0$,
we have recovered the temperature of non-accelerating BHs from
the Hawking temperature of the charged accelerating and rotating BHs \cite{[13]}.
Moreover, for $\beta=0$, $\ell=0$, $\tilde{k}=1$ and $\alpha=0$
in Eq.(\ref{aTH}), the Hawking temperature of the Kerr-Newman BH \cite{q6} is
recovered, which is reduced for $a=0$ to the temperature of the RN BH.
For $Q=0$, the temperature exactly reduces to the Hawking temperature
of the Schwarzschild BH at the residual mass of the BH \cite{[25]}.

In our analysis we have found that the quantum corrections decelerate the increase in temperature during the
radiation process. This correction causes the radiation ceased at some specific
temperature, leaving the remnant mass. The remnant will obtain at specific condition
\begin{equation}
M_{Res} \simeq \frac{M_{p}^2}{\beta_0\omega}\gtrsim \frac{M_{p}}{\beta_0}.\nonumber
\end{equation}
It is important to mention here that the value of the corrected Hawking temperature is smaller than
the original Hawking temperature and it stops radiating, when the mass of the BH reaches to its minimal
value $M_{Res}$. However, this result is still hold if background BH geometry is more generalized.

The results from the graphical analysis of corrected Hawking temperatures
with respect to the horizon for the given BH is summarized as
follows:
\begin{itemize}
\item For accelerating and rotating BH with $NUT$ parameter, the $T'_{H}$
decreases with the increasing horizon and BH reflects the stable state
for varying values of rotation parameter $a$ and correction parameter $\beta$.
The corrected temperature $T'_{H}$ also increases with the increase in $a$ and $\beta$.
The BH remnant can be obtained at nonzero horizon with maximum
temperature for different values of $\beta$ in the domain $0\leq r_{+}\leq15$.
\item The corrected temperature $T'_{H}$ decreases with the increase of $e$ and $g$.
For different values of electric and magnetic charges, the BH reflects stability
in the domain $0\leq r_{+}\leq15$.
\item The $T'_{H}$ increases with the increase in the value of arbitrary
parameter $k$.
\item The $T'_{H}$ increases with the increase in BH acceleration $\alpha$ and rotation parameter $\omega$.
\item In our analysis, we have considered the value $\Xi=0.01$, then the condition of GUP must be satisfied for
arbitrary values of $0\leq\beta<100$, the correction term is smaller
than the usual term as well as positive temperature is obtained. While, for $\beta>100$ the first order correction
term becomes greater than the usual term and the condition of GUP do not satisfy and
we observe the negative temperature which is non-physical. Furthermore,
for $\beta=100$, the semi-classical term cancel out with first order
correction term and hence the temperature vanishes.
\end{itemize}

\acknowledgements
A. \"{O}.~acknowledges
financial support provided under the Chilean FONDECYT Grant No. 3170035.

\end{document}